\documentclass[final,5p,twocolumn,times,sort&compress]{elsarticle}

\def\preprintdate{April 2025; 
published as Phys.\ Lett.\ B {\bf 867}, 139604 (2025)}

\def\al{\alpha}
\def\be{\beta}
\def\ga{\gamma}
\def\de{\delta}
\def\ep{\epsilon}

\def\et{\eta}
\def\th{\theta}

\def\la{\lambda}

\def\rh{\rho}

\def\si{\sigma}

\def\ta{\tau}

\def\ps{\psi}
\def\om{\omega}

\def\cl{{\cal L}}

\def\fr#1#2{{{#1}\over{#2}}}
\def\frac#1#2{{\textstyle{{#1}\over{#2}}}}
\def\tfrac#1#2{{\textstyle{{#1}\over {#2}}}}
\def\half{{\textstyle{1\over 2}}}
\def\quar{\frac 1 4}
\def\ol{\overline}
\def\prt{\partial}

\def\lsim{\mathrel{\rlap{\lower4pt\hbox{\hskip1pt$\sim$}}
    \raise1pt\hbox{$<$}}}
\def\gsim{\mathrel{\rlap{\lower4pt\hbox{\hskip1pt$\sim$}}
    \raise1pt\hbox{$>$}}}

\def\etal{{\it et al.}}

\newcommand{\beq}{\begin{equation}}
\newcommand{\eeq}{\end{equation}}
\newcommand{\bea}{\begin{eqnarray}}
\newcommand{\eea}{\end{eqnarray}}
\newcommand{\rf}[1]{(\ref{#1})}

\def\nn{\nonumber}

\def\psb{\ol\ps{}}

\def\cmtemplate#1#2#3#4{{#1}^{#3}_{#4}}

\def\bcm#1#2{\cmtemplate{b}{#1}{#2}{}}

\def\dcm#1#2{\cmtemplate{d}{#1}{#2}{}}

\def\gcm#1#2{\cmtemplate{g}{#1}{#2}{}}
\def\Hcm#1#2{\cmtemplate{H}{#1}{#2}{}}

\def\ctemplate#1#2#3#4{{#1}^{(#2)#3}_{#4}}

\def\bcf#1#2{\ctemplate{b}{#1}{#2}{F}}

\def\dcf#1#2{\ctemplate{d}{#1}{#2}{F}}

\def\gcf#1#2{\ctemplate{g}{#1}{#2}{F}}
\def\Hcf#1#2{\ctemplate{H}{#1}{#2}{F}}

\def\cmtemplate#1#2#3#4{{#1}^{#3}_{#4}}

\def\mn{{\mu\nu}}
\def\ma{{\mu\al}}
\def\mna{{\mu\nu\al}}
\def\ab{{\al\be}}
\def\bec{{\be\ga}}
\def\mab{{\mu\al\be}}
\def\mnab{{\mu\nu\al\be}}

\def\mabc{{\mu\al\be\ga}}
\def\mnabc{{\mu\nu\al\be\ga}}

\def\widecheck#1{\hskip#1pt\huge$\check{}$}
\def\bighacek#1#2{\vbox{\ialign{##\crcr\widecheck#2\crcr
  \noalign{\kern-9.5pt\nointerlineskip}
   $\hfil\displaystyle{#1}\hfil$\crcr}}}

\def\hb{\bighacek{b}{2}{}}

\def\hg{\bighacek{g}{3}{}}

\def\hH{\bighacek{H}{5}{}}

\begin{document}

\begin{frontmatter}

\title{
Relativistic spin precession in homogeneous background fields
}

\author{Yunhua Ding$^1$, V.\ Alan Kosteleck\'y$^2$, and Arnaldo J.\ Vargas$^3$}

\address{
$^1$Department of Physics and Astronomy, 
Ohio Wesleyan University, 
Delaware, OH 43015, USA\\
$^2$Physics Department, Indiana University, 
Bloomington, Indiana 47405, USA\\
$^3$Laboratory of Theoretical Physics, Department of Physics, 
University of Puerto Rico, 
R\'io Piedras, Puerto Rico 00936, USA
}

\address{}
\address{\rm 
\preprintdate}

\begin{abstract}
The Bargmann-Michel-Telegdi equation,
which describes the precession of the spin of a charged Dirac particle 
moving in a homogeneous electromagnetic field,
is generalized to include also other homogeneous background fields.
The treatment incorporates observable coefficients 
that govern operators of mass dimensions three through six 
in the underlying Dirac effective field theory.
A relativistic formulation valid in arbitrary inertial frames is obtained.
The results are applicable to searches for new physics
beyond the Standard Model,
including searches for Lorentz and CPT violation.
\end{abstract}

\end{frontmatter}

The precession of the spin of a Dirac particle
moving in a homogeneous electromagnetic field 
is described relativistically
by the Bargmann-Michel-Telegdi (BMT) equation~\cite{bmt59}.
The relativistic BMT equation is a classical equation
incorporating the Heisenberg equation of motion 
for the quantum-mechanical evolution of the spin operator,
and its 1959 formulation includes spin-precession effects 
from both an anomalous magnetic moment
and an electric dipole moment.
These effects involve the Larmor precession
from the magnetic-field torque on the magnetic moment,
the analogous precession for the electric dipole moment,
and the Thomas precession~\cite{lht26}
arising from the nonabelian nature of the Lorentz group 
and the particle acceleration.
The BMT equation applies in a wide variety of contexts
such as the motion of charged particles in storage rings
where it underlies,
for example,
experimental studies
of the anomalous magnetic moment of the muon~\cite{mu1,mu2,mu3}
and searches for electric dipole moments of particles
such as muons and protons~\cite{gb09,jp13,va16,rn19,yt22,ja22,sw23,edm1}. 

Our interest here lies in extending the relativistic BMT equation
to incorporate effects from homogeneous background fields
in addition to the electromagnetic field.
A homogeneous scalar background amounts to a mass term,
so in practice the backgrounds of interest carry Lorentz indices.
However,
a background with Lorentz indices
is a key characteristic of models with Lorentz violation,
which suggests the general extension of the BMT equation
can be framed in the context of Lorentz-violating theories.
These theories are motivated by the possibility of experimental detection 
of minuscule violations of the laws of relativity 
arising in an underlying unified theory such as strings~\cite{ks89,kp91}.
Effective field theory~\cite{sw}
is a powerful approach to describing small corrections
to established physics
and can be used to characterize Lorentz violation~\cite{kp95}. 
The general effective field theory for Lorentz violation
is constructed from Lorentz-violating operators
contracted with coefficients that determine the magnitudes
of the corresponding physical signals~\cite{ck97,ak04}.
The dominance of a given operator is associated with
its mass dimension $d$,
with lower $d$ governing leading-order effects
and values $d\leq4$ associated with operators that are renormalizable
in Minkowski spacetime. 
This framework is reviewed, for instance,
in Refs.~\cite{r1,r2,r3,r4,r5,r6},
and it provides the basis for numerous recent experimental studies 
of Lorentz violation~\cite{tables}.

In the present context,
we are interested in the spin precession of a Dirac particle
coupled to a prescribed electromagnetic field and other backgrounds.
Since the electromagnetic field is prescribed,
it could in principle be treated on the same footing 
as the other backgrounds,
but keeping it separate in the analysis permits 
distinguishing electromagnetic properties 
from prospective effects of new physics beyond the Standard Model.
The appropriate effective field theory for our purposes therefore involves 
a Lorentz-violating Dirac fermion coupled to an electromagnetic field.
For this system,
the operators affecting the behavior of the free particle
at arbitrary $d$ have been enumerated~\cite{ck97,km13},
and the construction of all electromagnetic couplings 
has been achieved~\cite{dk16,kl19}.
Here,
we focus on certain effects from additional homogeneous background fields
that provide dominant corrections to the relativistic BMT equation
involving the propagation and the electromagnetic interactions 
of the particle.
For propagation the values $d=3,4$ are included,  
while for interactions we consider operators with $d=5,6$
containing a factor of the electromagnetic field strength.

In the analysis that follows,
we disregard Lorentz-invariant but subdominant contributions
to the relativistic BMT equation.
Discussions of these effects can be found, 
for example, 
in Refs.~\cite{ak24,irf25} and references therein.
We work in Minkowski spacetime,
neglecting gravitational effects.
However,
corrections to the BMT equation 
arising from gravitational couplings 
that are invariant under local Lorentz transformations 
could in principle also be incorporated.
For a recent review of these corrections see, 
for example, 
Ref.~\cite{vnost23}.
We remark in passing that
the general local Lorentz and diffeomorphism violating
effective field theory for a Dirac particle
coupled to gravitational and electromagnetic fields
has been constructed~\cite{kl21}.
It offers a prospective path to establishing 
the corresponding gravitational modifications to the relativistic BMT equation 
via techniques analogous to those described here,
but this topic lies outside our present scope.

In what follows,
the particle is taken to have mass $m$ and charge $e$.
Its 3-velocity is denoted $\vec{\be}$
with corresponding Lorentz boost factor $\ga = 1/\sqrt{1-\be^2}$,
and its intrinsic 3-spin in the rest frame is denoted $\vec{s}$.
The particle is assumed to have a magnetic dipole moment given by 
$\vec{\mu} = (ge/2m)\vec{s}$,
where $g$ is the Land\'e factor related to the anomaly by $a = (g-2)/2$,
and an electric dipole moment given by
$\vec{d} = (\et e/2m)\vec{s}$.
The electric field $\vec{E}$ and the magnetic field $\vec{B}$
are taken to be homogeneous and externally prescribed.
We follow the notation and conventions of Ref.~\cite{ck97},
including the use of natural units with $c=\hbar=1$ 
and a Minkowski metric with negative signature.

The BMT equation can be expressed in a covariant manner
by introducing the particle 4-velocity $U^\mu$
and the intrinsic spin 4-vector $S^\mu$.
These quantities obey
\beq
U_\mu U^\mu = 1, 
\quad 
S_\mu S^\mu = -\tfrac{1}{4},
\quad
S_\mu U^\mu = 0, 
\label{conditions}
\eeq
where the first equation is the normalization condition
for the 4-velocity,
and the second equation sets the spin magnitude to 1/2.
The last equation expresses the orthogonality
of the 4-velocity and 4-spin,
which can be verified in the instantaneous rest frame
where $U^\mu = (1, \vec{0})$ and $S^\mu = (0, \vec{s})$.
Taking derivatives of the equations \rf{conditions}
with respect to the proper time yields the useful constraints 
\bea
U_\mu \fr{d U^\mu}{d \ta} = 0,
\quad
S_\mu \fr{d S^\mu}{d \ta} = 0,
\quad
U_\mu \fr{d S^\mu}{d \ta} = -S_\mu \fr{d U^\mu}{d \ta}.
\label{constraints}
\eea
Note that we are adopting here the formalism with $S^\mu$ 
used in the original BMT paper~\cite{bmt59,it29},
but the intrinsic spin could equivalently be described
via an antisymmetric 2-tensor dual to $S^\mu$ instead~\cite{jf26}.
The external homogeneous electromagnetic field 
is covariantly expressed in terms of an antisymmetric 2-tensor $F_{\mu\nu}$.
The electromagnetic field governs the 4-acceleration of the particle
via the usual relativistic equation incorporating the Lorentz force law,
\beq
\fr{d U^\mu}{d \ta} = \fr e m F^{\mn} U_\nu .
\label{force}
\eeq
The dynamical evolution of $S^\mu$ as a function of proper time $\ta$ 
is then determined by the relativistic BMT equation~\cite{bmt59},
\bea
\fr{d S^\mu}{d \ta}
&=&
\fr{e}{m} \Big[
(a+1) F^{\mn} S_\nu - a U^\mu F^{\nu\rh} U_\nu S_\rh
\nn\\
&&
\quad
-\fr{\et}{2} 
(\widetilde{F}^{\mn} S_\nu - U^\mu \widetilde{F}^{\nu\rh} U_\nu S_\rh)
\big],
\label{bmt}
\eea
where $\widetilde{F}^{\al\be} = \ep^{\al\be\mu\nu} F_{\mn}/2$
is the dual of $F_\mn$.

In a chosen inertial frame with time interval $dt = \ga d\ta$,
the BMT equation can be expanded to yield
the equation for the 3-spin as 
\beq
\fr{d\vec{s}}{dt} = \vec{\om}_{s}\times \vec{s},
\label{precession}
\eeq
where $\vec{\om}_{s}$ is the spin precession frequency.
The form of $\vec{\om}_{s}$ is
\bea
\vec{\om}_{s}
&=&
-\fr{e}{m}
\Big[
\big(a+\fr{1}{\ga}\big) \vec{B}
- \fr{a\ga}{\ga+1}(\vec{\be} \cdot \vec{B}) \vec{\be}
\nn\\
&&
\quad\quad 
-\big(a+\fr{1}{\ga+1}\big) \vec{\be} \times \vec{E} 
\nn\\
&&
\quad\quad 
+\fr{\et}{2} 
\big(\vec{E} -\fr{\ga}{\ga+1}(\vec{\be} \cdot \vec{E}) \vec{\be}
+\vec{\be} \times \vec{B}
\big) 
\Big].
\label{spinpf}
\eea
Together with the result~\rf{precession},
this expression is a generalization of the Thomas equation~\cite{lht27}
to include effects from an electric dipole moment.
These results are derived explicitly in Ref.~\cite{fs13},
encompassing other presentations 
in the literature~\cite{nspc59,ik98,mpr05,as05}.

One way to understand the physical content of the expression~\rf{spinpf}
is to note that the spin rate of change 
in the instantaneous particle rest frame is given by
\beq
\fr{d \vec{s}}{d \ta}\bigg|_{\rm {rest}} = 
-\fr{e}{2 m} (g \vec{B}^{\prime} +\et \vec{E}^{\prime}) \times \vec{s},
\label{dsdtrest}
\eeq
where contributions from both the magnetic and electric dipole moments
are included,
with $\vec{B}^{\prime}$ and $\vec{E}^{\prime}$ 
denoting the electromagnetic fields 
at the location of the particle in that frame.
The spin rate of change in an inertial laboratory frame is then given by
\beq
\fr{d \vec{s}}{d t}\bigg|_{\rm {lab}}
=\fr{1}{\ga}\fr{d \vec{s}}{d \ta}\bigg|_{\rm{rest}}
\hskip-5pt
+\vec{\om}_T \times \vec{s},
\label{rot}
\eeq
where 
\beq
\vec{\om}_T=\fr{\ga^2}{(\ga+1)} \fr {d \vec{\be}}{d t} \times \vec{\be}
\eeq
is the Thomas precession frequency~\cite{lht26,lht27}
accounting for the noninertial particle motion.
Converting the rest-frame fields $B^\prime$ and $E^\prime$ 
to laboratory-frame ones
using the standard boost transformations
\bea
\vec{B}^{\prime}
&=&
\ga(\vec{B}-\vec{\be} \times \vec{E})-\fr{\ga^2}{\ga+1} 
(\vec{\be} \cdot \vec{B}) \vec{\be} ,
\nn\\
\vec{E}^{\prime}
&=&
\ga(\vec{E}+\vec{\be} \times \vec{B})-\fr{\ga^2}{\ga+1} 
(\vec{\be} \cdot \vec{E}) \vec{\be} 
\label{betrans}
\eea
then recovers the expression~\rf{spinpf}.

Our goal is to extend the relativistic BMT equation~\rf{bmt}
to incorporate dominant contributions 
from the general effective field theory 
describing a Lorentz-violating Dirac fermion $\ps$
coupled to the electromagnetic field.
The Lagrange density $\cl_\ps$ of this theory 
contains the usual Dirac Lagrange density $\cl_0$
and a sum of terms $\cl^{(d)}$ containing operators of mass dimension $d$,
\beq
\cl_\ps =
\cl_0 
+ \cl^{(3)} 
+ \cl^{(4)} 
+ \cl^{(5)} 
+ \cl^{(6)} 
+ \ldots .
\label{nrlag}
\eeq
The explicit form of the free theory for arbitrary $d$
is given in Ref.~\cite{km13},
and the electromagnetic couplings are described in Ref.~\cite{kl19}.
We are interested here in leading-order corrections to the fermion propagator
that affect the BMT equation.
These involve spin-dependent operators 
at mass dimensions $d=3$ and $4$,
and they take the form~\cite{ck97}
\beq
\cl^{(3)} =
- \bcm 3 \mu \psb \ga_5 \ga_\mu \ps
- \half \Hcm 3 \mn \psb \si_\mn \ps
\label{lag3}
\eeq
and
\beq
\cl^{(4)} =
\half \dcm 4 {\ma} \psb \ga_5 \ga_\mu i\prt_\al \ps
+ \quar \gcm 4 {\mna} \psb \si_\mn i\prt_\al \ps
+ {\rm h.c.}
\label{lag4}
\eeq
Also of interest are corrections to the BMT equation
generated by leading-order interactions 
directly involving a factor of the homogeneous electromagnetic field $F_\mn$.
These arise from spin-dependent operators at mass dimensions $d=5$ and $6$
and are given by~\cite{dk16} 
\beq
\cl^{(5)}_F =
- \half \bcf 5 \mab F_\ab \psb \ga_5 \ga_\mu \ps
- \quar \Hcf 5 \mnab F_\ab \psb \si_\mn \ps
\label{l5f}
\eeq
and
\bea
\cl^{(6)}_F &=&
\quar \dcf 6 {\mabc} 
F_{\bec} \big( \psb \ga_5 \ga_\mu i\prt_{\al} \ps + {\rm h.c.} \big)
\nn\\
&&
+ \frac 1 8 \gcf 6 {\mnabc} 
F_{\bec} \big( \psb \si_\mn i\prt_{\al} \ps + {\rm h.c.} \big).
\label{l6f}
\eea
In these expressions,
the coefficients for Lorentz violation
$b^\mu$, $H^{\mn}$, $d^{\mn}$, $g^{\la\mn}$,
$\bcf 5 {\mu\ab}$,
$\Hcf 5 {\mn\ab}$,
$\dcf 6 {\mn\ab}$,
$\gcf 6 {\mn\rh\ab}$
are real and can be assumed to be constants
in an inertial frame in the solar system~\cite{ak04}.
Note that operators with an odd number of indices
are odd under the product CPT
of charge conjugation C, parity inversion P, and time reversal T,
and hence CPT violation in the theory
is controlled by coefficients with an odd number of indices.

Several techniques can be envisaged to derive
the desired extension of the relativistic BMT equation~\rf{bmt}.
One direct but comparatively lengthy method 
is to work at the level of relativistic quantum mechanics.
Using the Dirac equation obtained from $\cl$,
the 4$\times$4 hamiltonian $h$ for the system is constructed
and a Foldy-Wouthuysen transformation is obtained 
and used to convert $h$ to block-diagonal form.
The time derivative of the spin operator 
is found by commutation of the upper 2$\times$2 block of $h$ 
with the spin operator.
From this result,
the generalization of the spin precession frequency~\rf{spinpf}
that incorporates the effects of Lorentz violation is identified,
and the extended relativistic BMT equation is constructed.
This method has been used,
for example, 
to obtain the generalized spin precession frequency
for an antimuon trapped in a storage ring~\cite{bkl00,gkv14},
which has been adopted in experimental searches 
for Lorentz and CPT violation~\cite{lvmu1,lvmu2,lvmu3,lvmu4,lvmu5,lvmu6}.

In this work,
we adopt an alternative and computationally efficient technique 
to derive directly the desired leading-order contributions
to the relativistic BMT equation~\rf{bmt},
by taking advantage of symmetry properties 
of the Lorentz-violating operators and their coefficients.
The method consists of identifying
a complete set of independent contributions that 
are observer Lorentz covariant with appropriate parity,
are formed using coefficients for the operators with $d=3$ and $4$
in Eqs.~\rf{lag3} and~\rf{lag4},
and satisfy the conditions~\rf{constraints}.
The contributions from the operators with $d=5$ and $6$
in Eqs.~\rf{l5f} and~\rf{l6f}
can then be obtained using the field-redefinition protocol
established in Sec.~II~B~3 of Ref.~\cite{dk16},
which amounts to making the replacements
\bea
\bcm 3 \mu 
& \to &
\bcm 3 \mu 
+ \half \bcf 5 \mab F_\ab
,
\nn\\
\Hcm 3 \mn 
& \to &
\Hcm 3 \mn 
+ \half \Hcf 5 \mnab F_\ab
,
\nn\\
\dcm 4 \ma
& \to &
\dcm 4 \ma 
+ \half \dcf 6 \mabc F_{\be\ga}
,
\nn\\
\gcm 4 \mna 
& \to &
\gcm 4 \mna 
+ \half \gcf 6 \mnabc F_{\be\ga}.
\label{replacement}
\eea
This determines the desired extension of the relativistic BMT equation
up to an overall constant.
This constant can be fixed by comparing 
the Lorentz-violating contribution $(\om_{s, \rm{LV}})_3$
to the spin precession frequency
with the known result for a particle moving in a storage ring,
which in cartesian coordinates in the laboratory frame 
takes the form~\cite{bkl00,gkv14}
\bea
(\om_{s, \rm{LV}})_3 
&=&
-\fr{2}{\ga}(b_3-m_\mu g_3^{(\rm{A})})+2m_\mu d_{30}+2H_{12} 
\nn\\
&&
- \fr{2}{\ga}(1+\tfrac{3}{2} \be^2 \ga^2) m_\mu g_{120}^{(\rm{M})},
\label{muon3}
\eea
where the uniform magnetic field
is oriented in the $3$ direction perpendicular to the ring,
and where the superscripts $(A)$ and $(M)$ 
denote irreducible axial and mixed-symmetry combinations
of the coefficients $g_{\mn\al}$,
respectively~\cite{krt08,fr12}. 

To implement this technique,
note that we are seeking an expression for $dS^\mu/d\ta$ 
that is a sum of terms covariant under observer Lorentz transformations
and that must therefore be formed 
from the available covariant building blocks,
which are
$U^\mu$, $S^\mu$, $F_{\mn}$, the coefficients for Lorentz violation,
the metric $\et_\mn$, and the Levi-Civita tensor $\ep^{\mn\rh\si}$.
A given term in the sum can be taken linear in $S^\mu$ as usual,
satisfying the standard Uhlenbeck-Goudsmit hypothesis for spin~\cite{gu25}.
The restriction to leading-order effects
implies that the term can be assumed linear 
in $F_\mn$ and in the coefficients for Lorentz violation,
all of which are taken small compared to the particle mass
in the set of concordant reference frames
that are of experimental relevance~\cite{kl01}.
Investigating the alternative scenario with nonconcordant frames 
and large coefficients for Lorentz violation~\cite{klss24}
would be of definite interest but lies outside our present scope.

Both $F_\mn$ and the coefficients for Lorentz violation
are understood to be homogeneous or adiabatically changing
on the relevant experimental time scale,
so their derivatives can be neglected. 
Although some components of coefficients for Lorentz violation 
in the laboratory frame
are expected to vary as the Earth rotates and revolves around the Sun,
the time scale for these changes 
exceeds the typical measurement time scale,
such as that associated with a relativistic particle in a storage ring.
The contributions from the operators with $d=3$ or $4$ 
in Eqs.~\rf{lag3} and~\rf{lag4}
only affect propagation,
and hence the corresponding coefficients cannot appear combined with $F_\mn$
in any term in the extended BMT equation.
Similar reasoning shows that
contributions from a Lorentz-violating operator with $d=5$ or $6$ 
must involve coefficients appearing together with a single factor of $F_\mn$.
In effect,
both $F_\mn$ and the coefficients for Lorentz violation
can be understood as prescribed backgrounds,
which couple only in terms arising from interactions
in the underlying effective field theory.
We remark in passing that the various subleading contributions
from operators with other $d$ 
could also be investigated using the present technique.

Each term in the expression for $dS^\mu/d\ta$
must also be consistent with the discrete spacetime symmetries 
of the corresponding operator in the underlying effective field theory.
In establishing the explicit form of an allowed term,
it suffices to consider its parity,
as the conditions from time reversal and charge conjugation
then turn out to be satisfied automatically.
The extended BMT equation is linear in $S^\mu$,
so the parity of $S^\mu$ plays no role in the analysis.
The remaining factors of each term in the equation
must therefore be parity even.
This implies that the coefficients $b^\mu$ and $d^\mn$,
which are associated with operators containing the chiral matrix $\ga_5$,
must appear combined with a factor of $\ep^{\mn\rh\si}$.
 
The above procedure results in an expression for $dS^\mu/d\ta$,
which must then be subjected to the constraints~\rf{constraints}.
Verifying these constraints requires knowledge of 
the 4-acceleration $dU^\mu/d\ta$,
which in the Lorentz-invariant case is given by Eq.~\rf{force}.
In the present classical description of the spin dynamics,
this equation can be identified with the Heisenberg equation of motion
for the quantum wave packet associated with the particle.
It turns out that the homogeneity and perturbative nature
of the coefficients for Lorentz violation
ensure that no corrections arise to the 4-acceleration.
This result has been established in the literature in several ways.
One is to work with the classical lagrangian 
for the particle motion without an electromagnetic field,
noting that constant perturbative coefficients
guarantee conservation of the momentum
and hence conservation of the 4-velocity~\cite{kr10}.
Another approach is to take advantage of Finsler geometry~\cite{br,pf,bcs},
where perturbative coefficients for Lorentz violation
induce a classical particle to follow a geodesic
in a Finsler spacetime~\cite{ak04,ak11}.
Certain homogeneous coefficients are known to
generate special Finsler geometries 
known as Berwald spaces in which geodesics 
are conventional~\cite{bcs,ak11,mm74,hi75,ssay77,sk79,krt12,ek18},
thereby confirming the absence of corrections to Eq.~\rf{force}.
To ensure the extended BMT equation satisfies
the constraints~\rf{constraints}
involving $dS^\mu/d\ta$,
it therefore suffices to enforce the equations
\beq
S_\mu \fr{d S^\mu}{d \tau}\bigg|_{\rm LV} =0,
\quad
U_\mu \fr{d S^\mu}{d \ta}\bigg|_{\rm LV} = 0,
\label{LVconstraints}
\eeq
where LV denotes the restriction 
to Lorentz-violating terms in $d S^\mu/d \ta$.
Note that these conditions are observer Lorentz invariant
and thus hold in any inertial frame.

With the methodology in hand,
we can consider contributions to the extended BMT equation
from each type of Lorentz-violating operator in turn.
Consider first the $d=3$ operator in Eq.~\rf{lag3}
governed by the coefficient $b^\mu$.
We find that only one independent term satisfying the criteria 
for inclusion in the extended BMT equation can be constructed.
It can be written in the form 
\beq
\fr{d S^\mu}{d \ta}\bigg|_{\rm LV}
\supset
k_b \ep^{\mn\rh\si} b_\nu U_\rh S_\si,
\label{b}
\eeq
where $k_b$ is a constant.
Note that this expression satisfies the two conditions~\rf{LVconstraints} 
by virtue of the total antisymmetry of $\ep^{\mn\rh\si}$. 
To determine the constant $k_b$,
we evaluate the result~\rf{b} in the rest frame of the particle,
restrict attention to the spatial components,
and apply a boost to recover the result in the laboratory frame.
This yields the laboratory rate of change of the spin 
due to Lorentz violation as
\bea
\fr{d \vec{s}}{d t}\bigg|_{\rm {LV}}
\supset
\vec{\om}_{s, {\rm LV},b} \times \vec{s},
\eea
where
\beq
\vec{\om}_{s, {\rm LV},b} = 
\fr{k_b}{\ga} \vec{b}^{\prime} = 
\fr{k_b}{\ga}
\left(
\vec{b}+\fr{(\ga-1)}{\be^2}(\vec{\be} \cdot \vec{b}) \vec{\be}
-\ga \vec{\be} b_0 
\right),
\label{fb}
\eeq
with rest-frame spatial components of $b^\mu$ given by
$\vec{b}^{\prime}=(b^\prime_1, b^\prime_2, b^\prime_3)$ 
and laboratory-frame components of $b^\mu$ given by
$b^0$ and $\vec{b}=(b^1, b^2, b^3)$.
Matching the 3 component of the result~\rf{fb}
to the known expression~\rf{muon3} for a particle in a storage ring
requires averaging $\vec{\be}$ over the circular motion,
thereby revealing that $k_b=-2$. 
Finally,
the replacement~\rf{replacement} can be applied to Eq.~\rf{b}
to incorporate effects from the $b_F$-type coefficients.
We thus obtain the desired contribution
to the extended relativistic BMT equation as
\beq
\fr{d S^\mu}{d \ta}\bigg|_{\rm LV}
\supset
- 2\ep^{\mn\rh\si} b_\nu U_\rh S_\si
- \ep^{\mn\rh\si} b^{(5)}_{F\nu\ab} F^{\ab}U_\rh S_\si.
\label{bbf}
\eeq

Next,
we consider the $d=3$ operator in Eq.~\rf{lag3}
governed by the coefficient $H^\mn$.
Noting that $H^\mn = - H^{\nu\mu}$
and applying the same techniques as above,
we find that two independent terms 
satisfying the criteria can be constructed,
\beq
\fr{d S^\mu}{d \ta}\bigg|_{\rm LV}
\supset
k_{H1} H^{\mn} S_\nu+k_{H2} U^\mu H^{\nu\rh} U_\nu S_\rh,
\label{H}
\eeq
where $k_{H1}$ and $k_{H1}$ are constants.
The constraints~\rf{LVconstraints} are satisfied
provided $k_{H1}=-k_{H2}$.
Determining the spatial components of this equation
in the particle rest frame reveals a contribution 
to the spin precession frequency given by
\beq
\vec{\om}_{s, {\rm LV},H} = 
- \fr{k_{H1}} {\ga}
{\vec{H}_B^{\prime}},
\eeq
where
$\vec{H}_B^{\prime}=(-H^{\prime 23}, H^{\prime 13}, -H^{\prime 12})$
are coefficient components expressed in the rest frame.
Boosting to the laboratory frame yields
\beq
\vec{\om}_{s, {\rm LV},H} =
- k_{H1}\left(\vec{H}_B-\fr{\ga}{\ga+1}(\vec{\be} \cdot \vec{H}_B) \vec{\be}
-\vec{\be} \times \vec{H}_E\right),
\label{fH}
\eeq
with the rest-frame definitions 
$\vec{H}_B=(-H^{23}, H^{13}, -H^{12})$ 
and
$\vec{H}_E=(-H^{01}, -H^{02}, -H^{03})$.
Matching the 3 component of expression~\rf{fH}
to the result~\rf{muon3} for a particle in a storage ring
shows that $k_{H1}=-k_{H2}=2$. 
Implementing the substitution~\rf{replacement} in Eq.~\rf{H}
then yields the contribution  
\bea
\fr{d S^\mu}{d \ta}\bigg|_{\rm LV}
&\supset&
2 H^{\mn} S_\nu- 2 U^\mu H^{\nu\rh} U_\nu S_\rh
\nn\\
&&
+ \Hcf 5 {\mn\ab} F_{\ab} S_{\nu} - U^\mu \Hcf 5 {\nu\rh\ab} F_{\ab} U_{\nu} S_{\rh}
\nn\\
\label{hhf}
\eea
to the extended relativistic BMT equation.

Turning attention to the $d=4$ operators 
controlled by the coefficient $d^\mn$ in Eq.~\rf{lag4},
it is convenient to work in terms of 
the symmetric and antisymmetric combinations
$d^{(\mn)} = (d^\mn + d^{\nu\mu})/2$ 
and $d^{[\mn]} = (d^\mn - d^{\nu\mu})/2$,
which modulo a trace can serve as bases for irreducible representations
of the observer Lorentz group.
Examination of the possibilities reveals three independent contributions
satisfying the criteria for inclusion 
in the extended relativistic BMT equation,
\bea
\fr{d S^\mu}{d \ta}\bigg|_{\rm LV}
&\supset&
k_{d1} \ep^{\mn\rh\si} d_{[\nu\rh]} S_\si
+ k_{d2} U^\mu \ep^{\nu \rh \si \de} d_{[\nu \rh]} U_\si S_\de
\nn\\
&&
+ k_{d3} \ep^{\mn \rh \si} d_{(\nu \de)} S_\rh U_\si U^\de,
\label{d}
\eea
where $k_{d1}$, $k_{d2}$, and $k_{d3}$ are constants. 
The conditions~\rf{LVconstraints} are satisfied for $k_{d1}=-k_{d2}$.
Evaluating Eq.~\rf{d} in the rest frame
and extracting the spatial components
yields a contribution to the spin precession frequency of 
\bea
\vec{\om}_{s, {\rm LV},d} =
\fr{1}{\ga}(-2k_{d1}\vec{d}_{\rm A}^\prime+k_{d3}\vec{d}_{\rm S}^\prime),
\eea
with rest-frame coefficients 
$\vec{d}_A^\prime=-(d^{\prime [10]}, d^{\prime [20]}, d^{\prime [30]})$
and $\vec{d}_S^\prime=-(d^{\prime (10)}, d^{\prime (20)}, d^{\prime (30)})$.
Applying a boost transformation to the laboratory frame,
averaging $\vec{\be}$ over the orbit in the storage-ring, 
and comparing with the known result~\rf{muon3}
reveals that $k_{d1}=-m$ and $k_{d3}=2m$.
Together with the replacement~\rf{replacement} in Eq.~\rf{d},
this yields 
\bea
\fr{d S^\mu}{d \ta}\bigg|_{\rm LV}
&\supset&
-m \ep^{\mn\rh\si} d_{[\nu\rh]} S_\si
+ m U^\mu \ep^{\nu \rh \si \de} d_{[\nu \rh]} U_\si S_\de
\nn\\
&&
+ 2m \ep^{\mn \rh \si} d_{(\nu \de)} S_\rh U_\si U^\de
\nn\\
&&
-\frac{1}{2} m \ep^{\mn\rh\si} d^{(6)}_{F[\nu\rh]\ab} F^\ab S_\si
\nn\\
&&
+ \frac{1}{2} m U^\mu \ep^{\nu \rh \si \de} 
d^{(6)}_{F[\nu \rh]\ab} F^\ab U_\si S_\de
\nn\\
&&
+ m \ep^{\mn \rh \si} d^{(6)}_{F(\nu \de)\ab} F^\ab S_\rh U_\si U^\de
\label{ddf}
\eea
as the contribution from $d^{\mn}$
to the extended relativistic BMT equation.

Finally, 
we consider the $d=4$ operators
governed by the coefficient $g^{\mu \nu \rh}$ in Eq.~\rf{lag4}.
This coefficient can be decomposed 
into three irreducible components~\cite{krt08,fr12},
\bea
g^{\mu \nu \rh}=
g^{\mu \nu \rh(A)}+g^{\mu \nu \rh(T)}+g^{\mu \nu \rh(M)},
\eea
where the superscripts $(A)$, $(T)$, and $(M)$ 
denote axial, trace, and mixed-symmetry pieces,
respectively.
Following our methodology,
we find six independent terms obeying the requirements for inclusion 
in the extended relativistic BMT equation, 
\bea
\fr{d S^\mu}{d \ta}\bigg|_{\rm LV}
&\supset&
k_{g1} g^{\mn\rh(A)} S_\nu U_\rh
+k_{g2} g^{\mn\rh(T)} S_\nu U_\rh
\nn\\
&&
+k_{g3}  g^{\mn\rh(T)} U_\nu S_\rh
+k_{g4} g^{\mn\rh(M)} S_\nu U_\rh
\nn\\
&&
+k_{g5} g^{\mn\rh(M)} U_\nu S_\rh
+k_{g6} U^\mu g^{\nu\rh\si(M)} U_\nu S_\rh U_\si,
\nn\\
\label{girre}
\eea
where $k_{gj}$ with $j=1,2,3,...,6$ are constants.
To determine the values of these constants, 
we follow the same procedure as before.
Applying the two conditions~\rf{LVconstraints}
imposes $k_{g2}=k_{g3}=k_{g5}=0$ and $k_{g4}+k_{g6}=0$.
Working in the rest frame
and taking the $3$ component of Eq.~\rf{girre},
transforming it to the laboratory frame,
and comparing with the established result~\rf{muon3}
shows that $k_{g1}=k_{g4}=-2m$ and $k_{g6}=2m$.
Incorporating also the replacement~\rf{replacement} in Eq.~\rf{girre}
produces the desired contribution 
to the extended relativistic BMT equation in the form 
\bea
\fr{d S^\mu}{d \ta}\bigg|_{\rm LV}
&\supset&
-2m g^{\mn\rh(A)} S_\nu U_\rh
-2m g^{\mn\rh(M)} S_\nu U_\rh
\nn\\
&&
+2m U^\mu g^{\nu\rh\si(M)} U_\nu S_\rh U_\si,
\nn\\
&&
-m g^{(6)\mn\rh\ab(A)}F_\ab S_\nu U_\rh
\nn\\
&&
-m g^{(6)\mn\rh\ab(M)} F_\ab S_\nu U_\rh
\nn\\
&&
+m U^\mu g^{(6)\nu\rh\si\ab(M)} F_\ab U_\nu S_\rh U_\si.
\label{ggfirre}
\eea
Alternatively,
instead of working with irreducible components,
we can write an equivalent expression directly
in terms of the coefficient $g^{\mu \nu \rh}$,
\bea
\fr{d S^\mu}{d \ta}\bigg|_{\rm LV}
&\supset&
-2m g^{\mn\rh} S_\nu U_\rh
+2m U^\mu g^{\nu\rh\si} U_\nu S_\rh U_\si
\nn\\
&&
-m g_F^{(6)\mn\rh\ab}F_\ab S_\nu U_\rh
\nn\\
&&
+m U^\mu g_F^{(6)\nu\rh\si\ab}F_\ab U_\nu S_\rh U_\si.
\label{ggf}
\eea
In the rest frame,
the modifications to the spin precession frequency are found to be 
\bea
\label{fg}
\vec{\om}_{s, {\rm LV},g} =
\fr{2m}{\ga} \vec{g}^\prime ,
\eea
where 
$\vec{g}^\prime = (-g^{\prime 230}, g^{\prime 130},-g^{\prime 120})$.

At this stage,
we have in hand all the ingredients to generate
the complete form of the desired result.
Combining the usual relativistic BMT equation~\rf{bmt}
with the dominant 
contributions~\rf{fb},~\rf{hhf},~\rf{ddf}, and~\rf{ggfirre}
arising from the coefficients in the Lagrange 
densities~\rf{lag3},~\rf{lag4},~\rf{l5f}, and~\rf{l6f},
we obtain the full extended relativistic BMT equation as
\bea
\fr{d S^\mu}{d \ta}
&=&
\fr{e}{m} \left[
(a+1) F^{\mn} S_\nu - a U^\mu F^{\nu\rh} U_\nu S_\rh \right]
\nn\\
&&
-\fr{\et e}{2m} 
(\widetilde{F}^{\mn} S_\nu - U^\mu \widetilde{F}^{\nu\rh} U_\nu S_\rh)
\nn\\
&&
-2\ep^{\mn\rh\si} b_\nu U_\rh S_\si
+2 H^{\mn} S_\nu- 2 U^\mu H^{\nu\rh} U_\nu S_\rh
\nn\\
&&
-m \ep^{\mn\rh\si} d_{[\nu\rh]} S_\si
+m U^\mu \ep^{\nu \rh \si \de} d_{[\nu \rh]} U_\si S_\de
\nn\\
&&
+2m \ep^{\mn \rh \si} d_{(\nu \de)} S_\rh U_\si U^\de
\nn\\
&&
-2m g^{\mn\rh(A)} S_\nu U_\rh
-2m g^{\mn\rh(M)} S_\nu U_\rh
\nn\\
&&
+2m U^\mu g^{\nu\rh\si(M)} U_\nu S_\rh U_\si
\nn\\
&&
- \ep^{\mn\rh\si} b^{(5)}_{F\nu\ab} F^{\ab}U_\rh S_\si
+ \Hcf 5 {\mn\ab} F_{\ab} S_{\nu} 
\nn\\
&&
- U^\mu \Hcf 5 {\nu\rh\ab} F_{\ab} U_{\nu} S_{\rh}
\nn\\
&&
-\frac{1}{2} m \ep^{\mn\rh\si} d^{(6)}_{F[\nu\rh]\ab} F^\ab S_\si
\nn\\
&&
+\frac{1}{2} m U^\mu \ep^{\nu \rh \si \de} 
d^{(6)}_{F[\nu \rh]\ab} F^\ab U_\si S_\de
\nn\\
&&
+ m \ep^{\mn \rh \si} d^{(6)}_{F(\nu \de)\ab} F^\ab S_\rh U_\si U^\de
\nn\\
&&
-m g^{(6)\mn\rh\ab(A)}F_\ab S_\nu U_\rh
\nn\\
&&
-m g^{(6)\mn\rh\ab(M)} F_\ab S_\nu U_\rh
\nn\\
&&
+m U^\mu g^{(6)\nu\rh\si\ab(M)} F_\ab U_\nu S_\rh U_\si.
\label{bmtlv}
\eea
This equation describes the dominant effects on the spin precession 
of a Lorentz-violating Dirac particle 
with both a magnetic moment and an electric dipole moment
that is moving in a homogeneous electromagnetic field.
We note in passing that the equation also applies 
in the limit $e\to0$ of vanishing particle charge, 
as the terms~\rf{lag3},~\rf{lag4},~\rf{l5f}, and~\rf{l6f}
in the Lorentz-violating effective field theory
can {\it a priori} be taken independent of charge. 

We conclude this work with
an illustration of the application of the extended BMT equation~\rf{bmtlv}.
Consider a charged Dirac particle moving relativistically 
in a circular storage ring
that is designed to have a uniform magnetic field $\vec{B}$ 
perpendicular to the plane of the ring
and a radially constant electric field $\vec{E}$.
The electromagnetic 2-tensor $F^\mn$ in the laboratory frame
can then be written as
\beq
\label{fieldstrength}
F^{\mn}=\left(
\begin{array}{cccc}
0 & -E\cos\th & -E\sin\th & 0
\\
E\cos\th & 0 & -B & 0
\\
E\sin\th & B & 0 & 0
\\
0 & 0 & 0 & 0
\end{array}
\right),
\eeq
where the field magnitudes $B$ and $E$ are constant
and the angle $\th$ defines the position along the ring.
This idealized configuration is of relevance,
for example,
in searches for the electric dipole moment of charged particles 
such as muons and protons~\cite{gb09,jp13,va16,rn19,yt22,ja22,sw23,edm1}. 
In this scenario,
the fields $\vec{B}^\prime$ and $\vec{E}^\prime$ 
experienced by the particle in its instantaneous rest frame 
are constant in both direction and magnitude.
This can be seen by relating
the rest-frame fields $\vec{B}^\prime$ and $\vec{E}^\prime$ 
to the laboratory fields $\vec{B}$ and $\vec{E}$ 
using the boost transformations~\rf{betrans}.
Since $\vec{\be}$, $\vec{B}$, and $\vec{E}$
are an orthogonal set of vectors, 
$\vec{\be} \cdot \vec{B}=\vec{\be} \cdot \vec{E}=0$,
and hence
$\vec{B}^\prime$ is parallel to $\vec{B}$
and
$\vec{E}^\prime$ is parallel to $\vec{E}$.
The field magnitudes $B^\prime$ and $E^\prime$
are thus the same at all points around the ring.

Adopting this field configuration,
we seek the contribution $\vec{\om}_{s, \rm {LV}}$
from Lorentz violation to the spin precession frequency.
In the instantaneous particle rest frame,
the result~\rf{bmtlv} yields the expression
\bea
\vec{\om}_{s, \rm {LV}} =
-\fr{2}{\ga}(\vec{b}^{\prime} + \vec{H}_B^{\prime} 
- m \vec{d}_A^\prime - m \vec{d}_B^\prime - m \vec{g}^\prime),
\label{omrest}
\eea
where the primes indicate the rest-frame coefficients already introduced
with the replacements~\rf{replacement} implemented.
To convert this result to the laboratory frame,
a Lorentz boost is performed.
Assuming the particle moves 
counterclockwise at speed $\be$ around the storage ring,
the required instantaneous boost velocity is given by
$\vec{\be}=\be(-\sin\th, \cos\th, 0)$.
For definiteness,
we focus here on the spin precession frequency in the radial direction
and assume that the measurement process
on the relativistic particle
averages over the orbit around the ring.
The resulting boosted spin precession frequency~\rf{omrest}
must then be projected onto the radial direction 
by contracting it with the unit radial vector
$\hat{r}=(\cos\th, \sin\th, 0)$
and subsequently averaged over $\th$.
This procedure yields the contribution to the spin precession frequency
given by
\beq
\om^{\rm radial}_{s, \rm {LV}} =
\hH^{03}
+ E (\hb_{F}^{11} + \hb_{F}^{22} +\hg_{F}^{33})
+ B \hH_{F}^{33}
\label{omhacheck}
\eeq
in the laboratory frame.
In this expression,
we have introduced effective coefficients for Lorentz violation defined as
\bea
\hH^{03} &=&
-\beta[2 H^{0 3}-m(d^{1 2}-d^{2 1})
\nn\\
&&
\hskip 20pt
-\gamma m(2 g^{0 3 0}+g^{1 3 1}+g^{2 3 2})],
\nn\\
\hb_{F}^{11} &=&
\frac{1}{\gamma}b_F^{1 0 1}+H_F^{2 3 0 1}
 - md_F^{1 0 0 1} - \gamma mg_F^{2 3 0 0 1},
\nn\\
\hb_{F}^{22} &=&
+\frac{1}{\gamma}b_F^{2 0 2} - H_F^{1 3 0 2} 
 - m d_F^{2 0 0 2} + \gamma m g_F^{1 3 0 0 2},
\nn\\
\hg_{F}^{33} &=&
\gamma \beta^2 m (g_F^{0 3 1 0 2} -g_F^{0 3 2 0 1}),
\nn\\
\hH_{F}^{33} &=&
2\beta H_F^{0 3 1 2} - \beta m( d_F^{1 2 1 2} - d_F^{2 1 1 2}) 
\nn\\
&&
- \gamma \beta m( 2 g_F^{0 3 0 1 2} + g_F^{1 3 1 1 2} + g_F^{2 3 2 1 2} ).
\eea
These quantities have definite properties under rotation. 
Note that the structure of the indices appearing in result~\rf{omhacheck} 
correctly reflects the cylindrical symmetry of the storage ring configuration.
This symmetry also shows that the coefficients 
$\hb_{F}^{11}$ and $\hb_{F}^{22}$
must appear together and hence in this idealized scenario
cannot be measured independently.

\section*{Acknowledgments}

This work is supported in part 
by the U.S.\ Department of Energy under grant {DE}-SC0010120,
by a Teaching, Learning, and Innovation grant from Ohio Wesleyan University,
and by the Indiana University Center for Spacetime Symmetries.

\end{document}